\documentclass[fleqn,usenatbib]{mnras}

\usepackage{newtxtext,newtxmath}

\usepackage[T1]{fontenc}

\DeclareRobustCommand{\VAN}[3]{#2}
\let\VANthebibliography\thebibliography
\def\thebibliography{\DeclareRobustCommand{\VAN}[3]{##3}\VANthebibliography}

\usepackage{graphicx}	
\usepackage{amsmath}	

\usepackage{booktabs} 
\usepackage{threeparttable} 

\title[The Gum Nebula and a new distance technique]{Probing the morphology of the Gum Nebula through pulsar observables and a novel distance estimation method}

\author[Ashish et al.]{
Ashish Kumar,$^{1,2}$\thanks{E-mail: kalyanaastro@gmail.com}
Avinash A. Deshpande,$^{3}$
Pankaj Jain$^{3}$
\\
$^{1}$Department of Physics, Indian Institute of Technology, Kanpur-208016, India\\
$^{2}$National Centre for Radio Astrophysics, Tata Institute of Fundamental Research, Post Bag 3, Ganeshkhind, Pune 411007, India\\
$^{3}$Department of Space, Planetary $\&$ Astronomical Sciences $\&$ Engineering (SPASE), Indian Institute of Technology, Kanpur-208016, India}

\date{Accepted XXX. Received YYY; in original form ZZZ}

\pubyear{\the\year{}}

\begin{document}
\label{firstpage}
\pagerange{\pageref{firstpage}--\pageref{lastpage}}
\maketitle
\begin{abstract}
Various existing models of the Gum Nebula differ significantly in their parameters and suggested origins, which can be independently tested for consistency with data on some key observables of pulsars in the direction of the nebula. Our analysis of such data on the Vela pulsar, assuming a dominant scattering region in its foreground, suggests that the fractional distance of the scatterer is $0.89 \pm 0.01$, and for the given distance of the Vela pulsar, it translates to $254 \pm 16$ pc. Using independent distances of ten pulsars, we suggest a refined description of the Gum Nebula electron density model with its basic morphology similar to that used in the YMW16 model, which now provides better estimates of pulsar distances in these directions. In our new Gum Nebula model, as expected, the Vela pulsar would be behind the frontal edge of the Gum shell, which was intriguingly located in front of the nebula in the YMW16 model. We also present a new technique to better constrain the pulsar distances using their dispersion measure and temporal broadening simultaneously, and find that it is less affected by the uncertainties in the Galactic electron density distribution models. Notably, the new approach shows that the expected temporal broadening as a function of trial distance does not follow a monotonic increasing trend, but instead exhibits oscillations at regions of enhanced electron density. This behaviour is expected, as the method employs the integral form of temporal broadening with the appropriate weighting kernel, leading to more reliable estimates.
\end{abstract}
\section{Introduction}\label{sec:1}
\citet{gum1952large} discovered the largest optical structure in the southern hemisphere, a nebula with an angular extent of about 20$^\circ$, centred around G226$-$8. The nebula is possibly ionised by ultraviolet radiation from $\gamma^2$ Velorum\footnote{$\gamma$ Velorum is a quadruple system in the Vela constellation, consisting of two binary pairs: $\gamma^1$ (a blue giant and its fainter companion) and $\gamma^2$ Velorum (a Wolf--Rayet star and a blue supergiant).} and $\zeta$ Puppis\footnote{$\zeta$ Puppis is a massive O-type blue supergiant.}. Early spectral line studies have suggested that the Gum Nebula is an evolved Str\"omgren sphere powered by these two stars, although with differing estimates of the nebular parameters \citep{beuermann1973reexamination, reynolds1976observations, chanot1983gum}. Alternative formation channels of the nebula have also been put forward. \citet{brandt1971gum} and \citet{alexander1971gum} argued that the nebula is extremely hot ($\gtrsim 54{,}000$\,K), which requires a non-stellar source of excitation. They proposed that the Vela supernova had created the nebula, and described it as a ``fossil Str\"omgren sphere'', a relic ionised shell now cooling and recombining since the ionisation source is no longer present. Other studies instead attributed the nebula's origin to a supernova explosion (either Vela or the former companion of $\zeta$ Puppis), with its current ionisation provided by $\gamma^2$ Velorum and $\zeta$ Puppis \citep{reynolds1976gum, woermann2001kinematics}. \citet{reynolds1976gum} proposed that the stellar wind from $\zeta$ Puppis could be sufficient to ionise the nebula, an idea later supported by theoretical work on wind-blown bubbles by \citet{weaver1977interstellar}. \citet{reynoso1997cold} suggested that multiple supernova explosions could also account for the nebula, compressing the gas into giant shells with a radius of hundreds of parsecs. The nebular parameters suggested by the aforementioned studies are consistent with their respective proposed origin.

The Gum Nebula exhibits a complex structure, encompassing numerous cometary globules \citep{zealey1983} as well as the prominent IRAS Vela Shell (IVS; \citealt{sahu1993kinematics, gao2025}). The physical association of the IVS with the Gum Nebula has been debated for decades, but recent 3D dust-mapping by \citet{gao2025} has established that the IVS is indeed an integral component of the nebula. Their analysis, combining Gaia and Hipparcos astrometry, further indicates that the massive stars $\gamma^2$ Velorum and $\zeta$ Puppis lie within the IVS and are the dominant sources driving its ionisation. \citet{purcell2015radio} have studied the Gum Nebula region using the radio polarisation data from the VLA and the S-band Parkes All-Sky Survey, together with H\,$\alpha$ images from the Southern H-Alpha Sky Survey Atlas, and have provided the updated physical parameters and properties of the Gum Nebula. They have modelled the nebula as a spherical shell of uniform electron density and have obtained an average electron density of $1.3 \pm 0.4$ cm$^{-3}$, an angular radius of $22.7^\circ \pm 0.1^\circ$, a shell thickness of $18.5^{+1.5}_{-1.4}$ pc, and an ambient magnetic field strength of $3.9^{+4.9}_{-2.2}$ $\mu$G.

Pulsar observables provide a tool to study the interstellar medium (ISM) and such discrete nebular structures. The ionised ISM introduces a frequency-dependent delay to the pulsar signals, which is characterised by the dispersion measure (DM). It represents the electron column density between the observer and the pulsar (${\rm DM} = \int_0^D n_e dz$). In addition to dispersion, the inhomogeneous ISM scatters the pulsar signals, leading to multipath propagation that gives rise to scintillation and other observable effects, like pulse broadening, angular broadening, etc. Particularly, the studies of the pulsars in or behind the Gum Nebula region have shown enhanced DM and scattering, as well as variations in scattering across the nebula \citep{mitra2001}. Noting this, almost all of the Galactic Electron Density Models (GEDMs) include explicit contributions to the electron density distribution associated with the Gum Nebula. Many of these Gum Nebula Electron Density (GNED) models are motivated by some of the aforementioned mechanisms of the nebula origin. The correctness of these models and the parameters can be assessed effectively by using data on pulsars towards the nebula. More specifically, the model prediction of the DM-based distances and scatter broadening can be compared with those measured. Fortunately, independent distance estimates are available for some of the pulsars in these directions. The independent distance estimates are typically derived from parallax measurements \citep{deller2019, rioja2020, kumar2025}, associations with globular clusters or supernova remnants (SNRs; \citealt{kulkarni1993}), and kinematic distances based on 21 cm H\,{\sc i} absorption in pulsar spectra combined with Galactic rotation models \citep{johnston1996}. 

In several directions across the sky, the ISM exhibits enhanced electron densities that are concentrated in localised regions, such as the Gum Nebula shell. Such regions along the line of sight invariably manifest as discrete scattering screens or dominant scatterers. In specific cases, it is possible to constrain the relative distance of a discrete dominant scatterer for a pulsar, if any, utilising a technique introduced by \cite{deshpande1998improving}, where estimates of certain required scintillation parameters and proper motion are available. Using such data available for the Vela pulsar and its precise parallax-based distance \citep{dodson2003vela}, it has now become possible to estimate the distance to the dominant scatterer for the Vela pulsar. Since the Vela pulsar is believed to be a part of the Gum Nebula, as suggested by several studies \citep{backer1974, kirsten2019}, and therefore, the distance to the frontal edge of the nebula can be inferred. This, in turn, provides constraints on the size of the nebula relevant to the DM and scintillation parameters of the pulsars towards the nebula. Further, the comparison of independently determined distances for ten pulsars with those predicted by GEDMs (based on DM) has provided a critical assessment of the Gum Nebula contribution in the respective models. The nebula component in the YMW16 model, in comparison with other models, provides better consistency between predicted and independently estimated distances for pulsars towards this region. However, we have found that significant distance improvement is possible due to the refinement of parameters characterising nebula contribution.

In the following Section \ref{sec:2} describes the details of our analysis of the data on the Vela pulsar to constrain the distance of the frontal edge of the Gum Nebula. Section \ref{sec:3} presents the comparison of GEDMs and our improved model parameters for the Gum Nebula. A new method is explored to better constrain the pulsar distances using their DM and temporal broadening, and is presented in Section \ref{sec:4}. Section \ref{sec:5} contains a summary of our results, discussion, and conclusion.

\section{Frontal Edge of the Gum Nebula}\label{sec:2}
In view of the range of suggestions for the Gum Nebula parameters, we explore the possibility of using the Vela pulsar observables to independently constrain the distance to the scattering screen rendered by the Gum Nebula shell. More specifically, to estimate the distance to the frontal edge of the Gum Nebula, we have used the \cite{deshpande1998improving} technique, which has been demonstrated to estimate the distance to the pulsar. This technique has been motivated by the work of \cite{gwinn1993angular}, where they assume a two-component model for the scattering medium consisting of a uniform component plus a thin discrete scattering screen. The technique uses the measurements of angular broadening, temporal broadening, scintillation timescale, and proper motion of pulsars to estimate both the location of the dominant scattering screen and the scattering strength factor (ratio of the mean scattering rate in a discrete thin screen to uniform distribution) along the line of sight. The angular broadening ($\theta_{\rm H}$; \citealt{alcock1978, blandford1985}) and temporal broadening ($\tau_{\rm sc}$; \citealt{blandford1985}) for a source at distance $D$ are given by 
\begin{equation}
    \theta^2 = \dfrac{1}{D^2} \int_0^D z^2 \psi(z) dz,
    \label{eq:ang_broad}
\end{equation}
\begin{equation}
    \tau_{\rm sc} = \dfrac{1}{2cD} \int_0^D z(D-z) \psi(z) dz,
    \label{eq:temp_broad}
\end{equation}
where $\psi(z)$ represents the mean scattering rate per unit length, and $c$ is the speed of light. \cite{gwinn1993angular} have utilised the ratio of the observed angular broadening ($\theta_{\rm H}$) to that estimated from the measured temporal broadening:
\[
\theta_\tau = \sqrt{4 \ln(2) \left( \frac{4 c \tau_{\rm sc}}{D} \right)}.
\]
\cite{deshpande1998improving} extended this approach by also incorporating the ratio of the measured transverse proper motion velocity ($V_{\rm PM}$) to that inferred from temporal broadening and the diffractive scintillation timescale ($t_{\rm diff}$; ratio of the coherence length scale to effective traverse velocity of the pulsar):
\[
v_{\rm iss}^2 = \frac{D c}{4 \pi^2 \tau_{\rm sc} t_{\rm diff}^2 \nu^2},
\]
where $\nu$ is the observing frequency.
By solving equations (\ref{eq:ang_broad}) and (\ref{eq:temp_broad}) and using the above-mentioned ratios (angular broadening values ratio, $r_\theta$ and transverse velocities ratio, $r_v$), \cite{deshpande1998improving} derived the following expression to estimate the pulsar distance:
\begin{equation}
    D^2 r_\theta^2 r_v^2 (2x - 3x^2) + D r_v^2 (2x - 1) + (3x - 1)(x - 1) = 0,
    \label{eq:1}
\end{equation}
where $x = d_s / D$ is the fractional distance to the dominant scatterer, and $d_s$ is the physical distance of the scatterer.

However, the method requires knowledge of the distance to the dominant scatterer to translate the fractional scatterer distance to the physical distance of the pulsar. This technique can be inverted to estimate the distance to the dominant scatterer in the line of sight of the pulsar whose independent distance is known. We apply this technique to the Vela pulsar to estimate the distance to the frontal edge of the Gum Nebula. The required aforementioned observables of the Vela pulsar are listed in Table \ref{table:vela_para}. \cite{krishnakumar2015scatter} have provided the refined temporal broadening value for the Vela pulsar at 327 MHz, which agree within $1\,\sigma$\footnote{The sigma is the quadratic sum of both measurements uncertainties, the same convention is used throughout the paper} with that of the \cite{desai1992speckle} estimates at 2.3 GHz, assuming a spectral index of $\alpha = -4$ in the scaling relation $\tau_\text{sc} \propto \nu^{\alpha}$ \citep{romani1986refractive} and the relation $2\pi\Delta\nu_\text{sc} \tau_\text{sc} \approx 1$ \citep{cordes1998}. However, it is worth noting that these two temporal broadening estimates would deviate from each other by $\sim 8\,\sigma$ if $\alpha$ is to be -4.4. To avoid such frequency dependency, we adopt the temporal broadening estimates of \cite{desai1992speckle} in the analysis, since the same study also provides the other two required observables, $\theta_\text{H}$, $t_{\text{diff}}$, at the same frequency.
\begin{table*}
    \centering
    \resizebox{\textwidth}{!}{%
    \begin{threeparttable}
    \begin{tabular}{cccccccccccc}
    \toprule
    \addlinespace
    Pulsar & R.A.\tnote{a} & Dec.\tnote{a} & $\mu_\alpha$\tnote{a} & $\mu_\delta$\tnote{a} & $D$\tnote{a} & $\Delta\nu_\text{sc}$\tnote{b} & $\tau_\text{diff}$\tnote{b} & $\theta_\text{H}$\tnote{b} \\
    \addlinespace
     & (hms) & (dms) & (mas yr$^{-1}$) & (mas yr$^{-1}$) & (pc) & (kHz) & (s) & (mas)  \\
    \midrule
    \addlinespace
    J0835-4510 & 08:35:20.61149(2) & -45:10:34.8751(3) & $-49.68 \pm 0.06$ & $29.9 \pm 0.1$ & 287$^{+19}_{-17}$ & $68 \pm 5$ & 15 & $1.6 \pm 0.2$\\
    \bottomrule
    \end{tabular}
    \begin{tablenotes}
        \item(a) \cite{dodson2003vela}
        \item(b) \cite{desai1992speckle}; $\Delta\nu_{\rm sc}$ represents the half-width at half-maximum (HWHM) of the autocorrelation function of the pulsar dynamic spectrum along the frequency axis ($\tau = 0$), $t_{\rm diff}$ is the half-width at the $1/e$ power level along the time axis ($\nu = 0$), and $\theta_{\rm H}$ denotes the full width at half maximum (FWHM) of the scattering disk size.
    \end{tablenotes}
    \caption{List of observable parameters of the Vela pulsar.}
    \label{table:vela_para}
    \end{threeparttable}
    }
\end{table*}

\begin{figure*}
    \centering
    \includegraphics[width=0.48\linewidth]{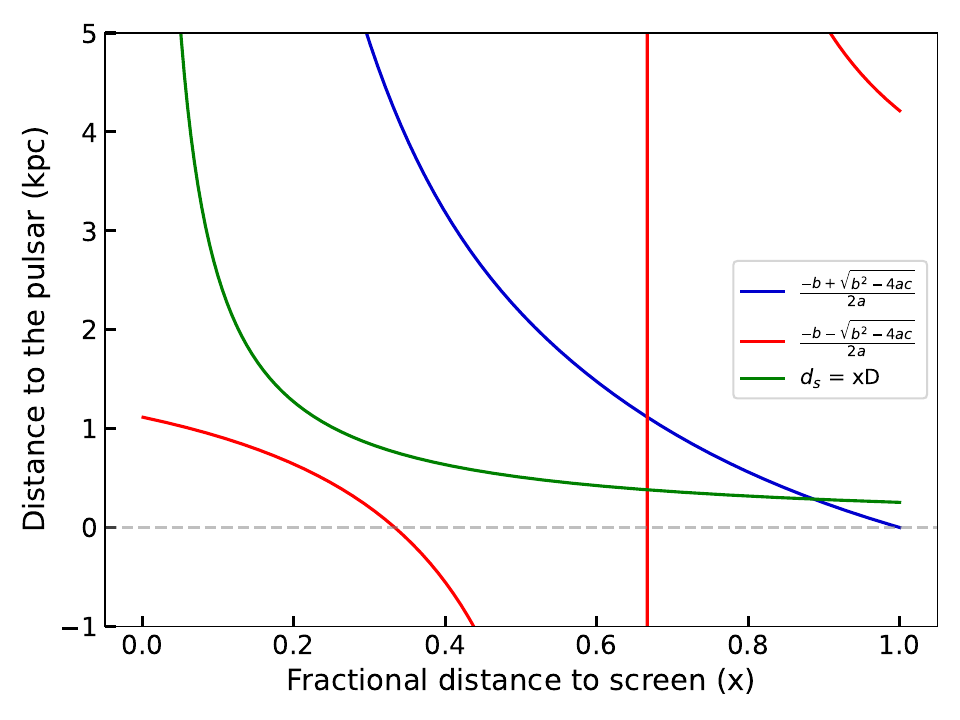}
    \includegraphics[width=0.51\linewidth]{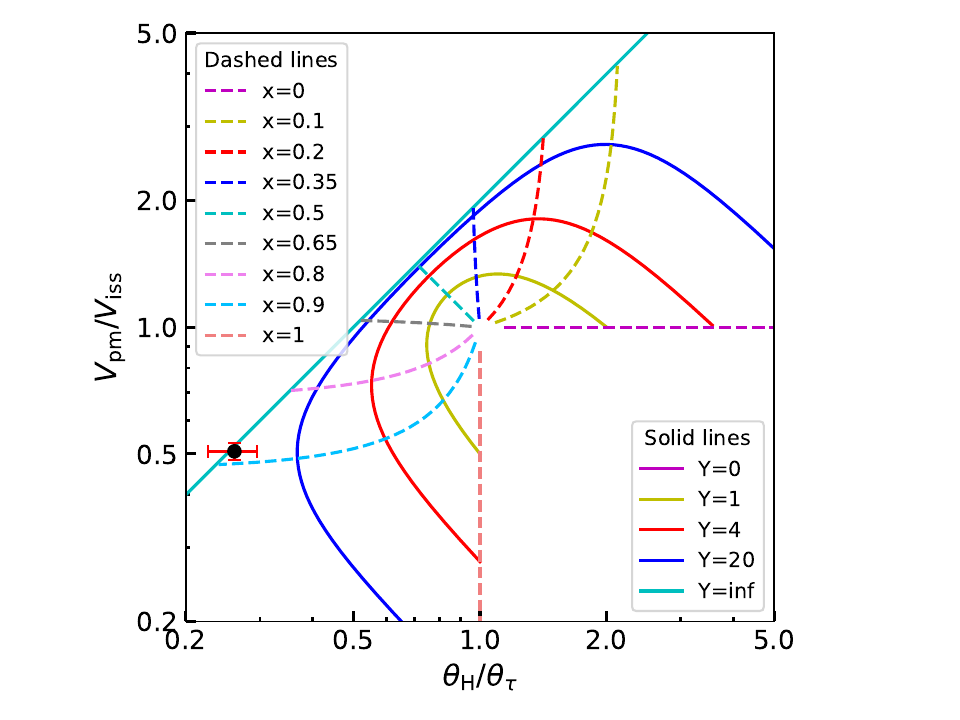}
    \caption{Left: Both roots of the quadratic equation (\ref{eq:1}) are plotted against the fractional scatterer distance, shown in blue and red. The green curve represents $D\ =\ d_s/x$, where the scatterer distance is 254 pc. The physical solution for the fractional scatterer distance corresponds to the intersection of blue and green curves at $x$ = 0.89. The red curve reaches unrealistic distance estimates for a range of fractional scatterer distance values. Right: The variation of $R_\theta$ and $R_\text{v}$ with the fractional scatterer distance and scattering strength factor ($Y$) is shown. The black points with error bars indicate the estimated values for the Vela pulsar, with $r_{\theta}$ = 0.49 kpc$^{-1/2}$, $r_\text{v}$ = 0.95 kpc$^{-1/2}$. The estimated fractional scatterer distance and scattering strength factor for the Vela pulsar are $0.89 \pm 0.01$, 384, respectively.}
    \label{fig:3}
\end{figure*}

Using parameter values listed in Table \ref{table:vela_para} and equation (\ref{eq:1}), we have estimated the fractional scatterer distance to be $0.89 \pm 0.01$, which translates to the scatterer distance of $254 \pm 16$ pc for the given Vela pulsar distance of $287^{+19}_{-17}$ \citep{dodson2003vela}. The other root of the quadratic equation (\ref{eq:1}) translates to unrealistic distance estimates, as can be seen in Fig. \ref{fig:3}. The uncertainty in fractional scatterer distance ($x$) is computed using a Monte Carlo approach with 100,000 samples. For each contributing parameter ($D$, $\theta_\text{H}$, $\tau_\text{sc}$, $t_{\text{diff}}$, and proper motion), random samples have been drawn from normal distributions that have a mean value and standard deviation equal to those in the respective parameters. For each sample, the fractional scatterer distance is computed, and the $1\,\sigma$ error bar is determined using the 16$^{\rm th}$ and 84$^{\rm th}$ percentiles values of the distribution. Our estimated fractional scatterer distance differs by approximately $2\,\sigma$ from the value reported by \cite{desai1992speckle}, which likely arises from their assumed Vela pulsar distance of $500 \pm 100$ pc \citep{frail1990}. 

The Vela pulsar is well known to be associated with the Gum Nebula, whose frontal edge is most likely the dominant contributor to the Vela pulsar scattering \citep{backer1974, kirsten2019, wang2025}. If indeed so, our estimated distance to the dominant scatterer for the Vela pulsar implies the Gum Nebula radius of $196 \pm 16$ pc, corresponding to an angular radius of $24^\circ \pm 2^\circ$, which agrees within $1\,\sigma$ of the \cite{purcell2015radio} estimate. Here, as per \cite{purcell2015radio} study, we have assumed that the Gum Nebula is spherical, centred at ($l$ = 261$^\circ$, $b$ = -2.5$^\circ$; \citealt{woermann2001kinematics}) with a distance of 450 pc. \citet{wang2025} have estimated the distance to the dominant scatterer for PSR B0740–28 to be $245^{+69}_{-72}$\,pc, based on measurements of interstellar scintillation speed. They assumed the Gum Nebula is centred on G258-2 at 350\,pc, with its frontal edge at approximately 240\,pc, which is consistent with their inferred scatterer distance for B0740-28. Using the same Gum Nebula parameters (3D centre coordinates), the scatterer distance for the Vela pulsar places the nebula’s frontal edge at $250 \pm 15$\,pc, which agrees within the uncertainties with their estimate of frontal edge distance.

For completeness, we have checked if any imprint of the Gum Nebula is apparent in the Gaia data by examining Gaia sources in the $80^\circ\ \times\ 80^\circ$ region around the nebula centre, such as any source density enhancement in the Gum Nebula shell. However, no such enhancement is apparent, and in fact, compared to the centre, the source density slightly falls towards the shell. Although any possible causal connection between this fall and the presence of the Gum Nebula shell is unclear, the delineation of the boundary of this region (as well as the assumed centre) is consistent with the scale of this fall in Gaia source density.

Several studies have shown that H\,{\sc ii} regions along pulsar sight lines can significantly affect observables such as dispersion, scattering, and rotation measures, and have even been used to identify and study such regions directly \citep{mitra2003, harvey2011, ocker2024}. In particular, scattering due to discrete H\,{\sc ii} regions has been explicitly assessed by several studies \citep{ocker2020, mall2022, ocker2024a}. Motivated by these studies, we examined Gaia sources within a $20^\circ \times 20^\circ$ region centred on the Vela pulsar to search for hot O/B-type stars whose Str\"omgren spheres might contribute to discrete density enhancements \citep{prentice1969h}. No such stars were found in close proximity to the line of sight, reinforcing the conclusion that the frontal edge of the Gum Nebula itself is the primary scattering screen for the Vela pulsar.

\section{GEDMs Comparison and Refinements}\label{sec:3}
GEDMs are primarily constructed from our understanding of the electron density distribution of our galaxy, which can be from pulsar DM, spectral line observations, H\,$\alpha$ surveys, etc. However, accurately measuring the parameters describing these models remains a challenge. Observables of the pulsars play a key role in refining and constraining GEDMs. Two of the most widely used GEDMs till now are NE2001 \citep{cordes_ne2001} and YMW16 \citep{yao2017new}, which are essential for estimating the distances of pulsars and fast radio bursts (FRBs) and for describing the properties of the interstellar medium (ISM). These models include the Gum Nebula as a separate electron density-enhanced region. The component(s) of the Gum Nebula used by these models and their respective parameters are shown in Table \ref{table:gn_components}. The NE2001 model assumes the nebula to be a spherical region with an angular radius of $\sim 30^\circ$. It also includes two electron density enhanced spherical regions (clumps) and one electron deficit region (void) to match the measured pulsar observables to those predicted by the model in these directions; for more details, see \cite{cordes_ne2001}. While based on the current understanding of the Gum Nebula \citep{purcell2015radio, woermann2001kinematics}, \cite{yao2017new} have modelled nebula as an ellipsoidal shell,
\begin{equation}\label{eq:ellipse}
    \Big(\dfrac{x - x_{\rm GN}}{A_{\rm GN}}\Big)^2 + \Big(\dfrac{y - y_{\rm GN}}{A_{\rm GN}}\Big)^2 + \Big(\dfrac{z - z_{\rm GN}}{K_{\rm GN} A_{\rm GN}}\Big)^2 = 1,
\end{equation}
where ($x_{\rm GN},\ y_{\rm GN},\ z_{\rm GN}$) are the galactocentric Cartesian coordinates of the nebula centre\footnote{It is a right-handed coordinate system centred at the Galactic Centre, with the $x$-axis parallel with $l\ =\ 90^\circ$ and the $y$-axis with $l\ =\ 180^\circ$, where $(l,\ b)$ represent the standard Sun-centred Galactic coordinates. The Sun is located at ($x\ =\ 0,\ y\ =\ 8300\ {\rm pc},\ z\ =\ 6.2\ {\rm pc}$).
}, 
$K_{\rm GN}$ is the ratio of the $z$ axis dimension to that of the $x-y$ plane, and $A_{\rm GN}$ is the radius of the shell in the $x-y$ plane. The assumed excess electron density in the shell has a Gaussian profile with 1/e half-width $W_{\rm GN}$:
\begin{equation}\label{eq:gn_ne}
n_{\rm GN} = n_{\rm GN_0}\exp\left[-\left(\frac{s_{\rm GN}}{W_{\rm GN}}\right)^2\right],
\end{equation}
where $s_{\rm GN}$ is the perpendicular distance to the mid-point of the ellipsoidal shell (for more details, see \citealt{yao2017new}).

\begin{figure}
    \centering
    \includegraphics[width=1\linewidth]{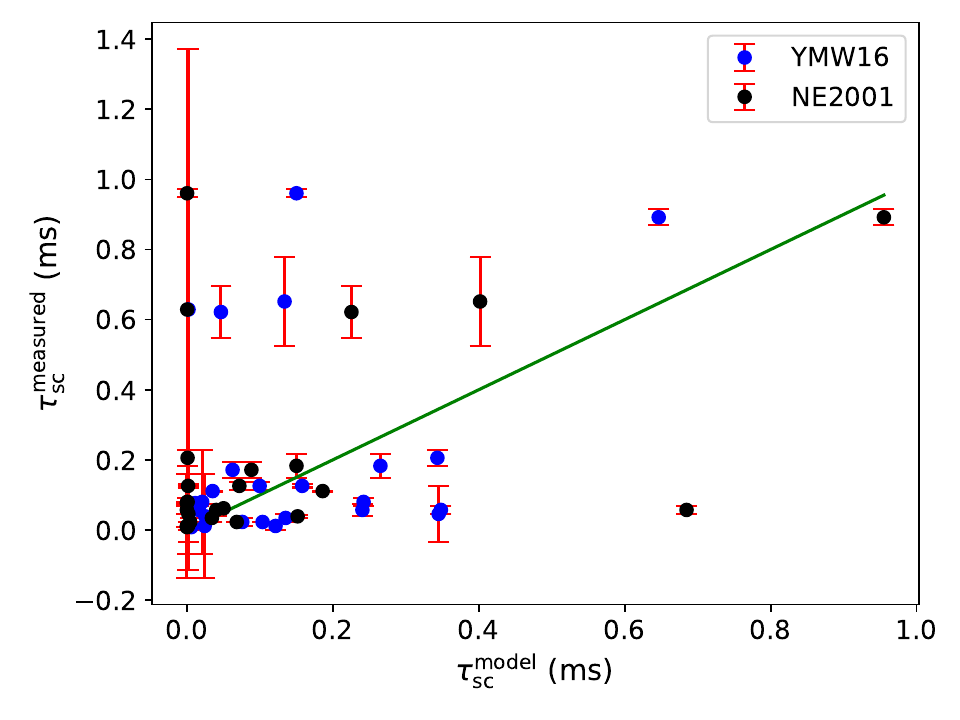}
    \caption{The plot shows a comparison between the model-predicted and measured temporal broadening values of the pulsars in the direction of the Gum Nebula at 1 GHz. The measured temporal broadening values are scaled to 1 GHz using the frequency scaling spectral index, $\alpha$ = -4. The values on or close to the green line ($y = x$) have less deviation from the model predictions.}
    \label{fig:tsc_comparison}
\end{figure}

The reliability of these models can be assessed by comparing model-predicted and independently measured pulsar observables \citep{Price_Flynn_Deller_2021, yao2017new}. Comparing DM-based distances with independent estimates of distances, we have examined the Gum Nebula electron density models within these GEDMs. The temporal broadening values of the pulsars are not considered in the analysis since the model-predicted temporal broadening values show a huge deviation from that of the measured values (Fig. \ref{fig:tsc_comparison}), which is a clear indication that the DM$^{2.2}$ model \citep{rmd97, krishnakumar2015scatter} is not a good estimator for temporal broadening of pulsars in these directions, as well as in general, the temporal broadening does not simply follow the DM$^\beta$ ($\beta>1$) but also depends on the fractional distances to the scatterers (as discussed in Section \ref{sec:4}).

\begin{figure}
    \centering
    \includegraphics[width=1\linewidth]{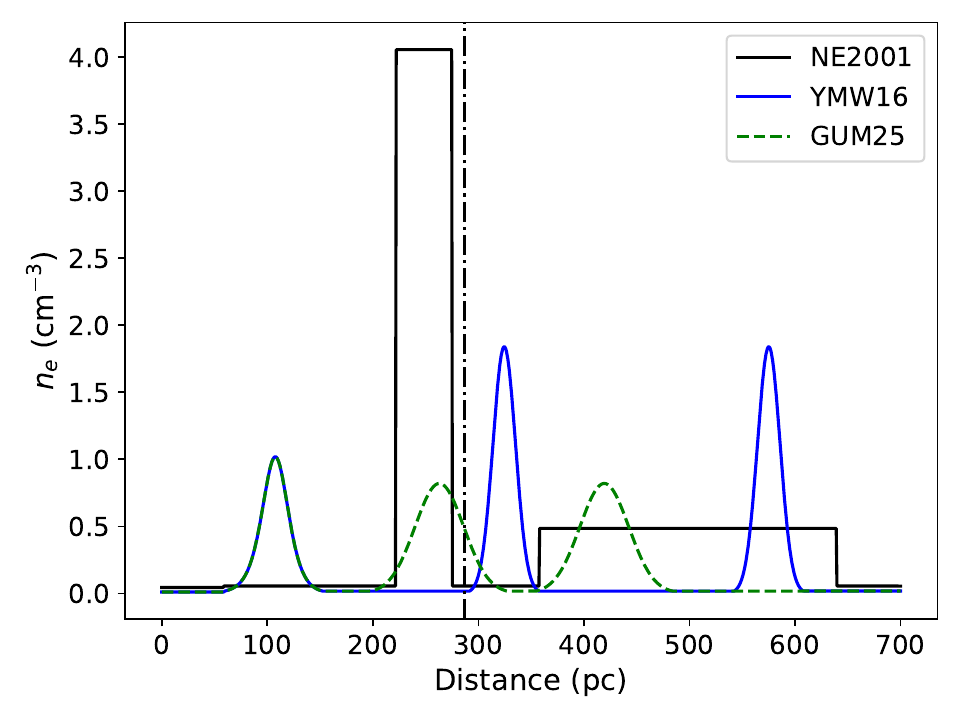}
    \caption{The plot illustrates the electron density profile along the line of sight to the Vela pulsar. The dashed-dotted vertical line marks the location of the Vela pulsar. The first peak, around 100\,pc, corresponds to the electron density enhancement in the Local Bubble, followed by two peaks associated with the Gum Nebula. In the YMW16 model, the Vela pulsar appears in the foreground of the Gum Nebula shell, whereas in the NE2001 and modified model (GUM25), it is positioned just behind the frontal edge of the nebula.}
    \label{fig:vela_ne}
\end{figure}

\begin{figure}
    \centering
    \includegraphics[width=1\linewidth]{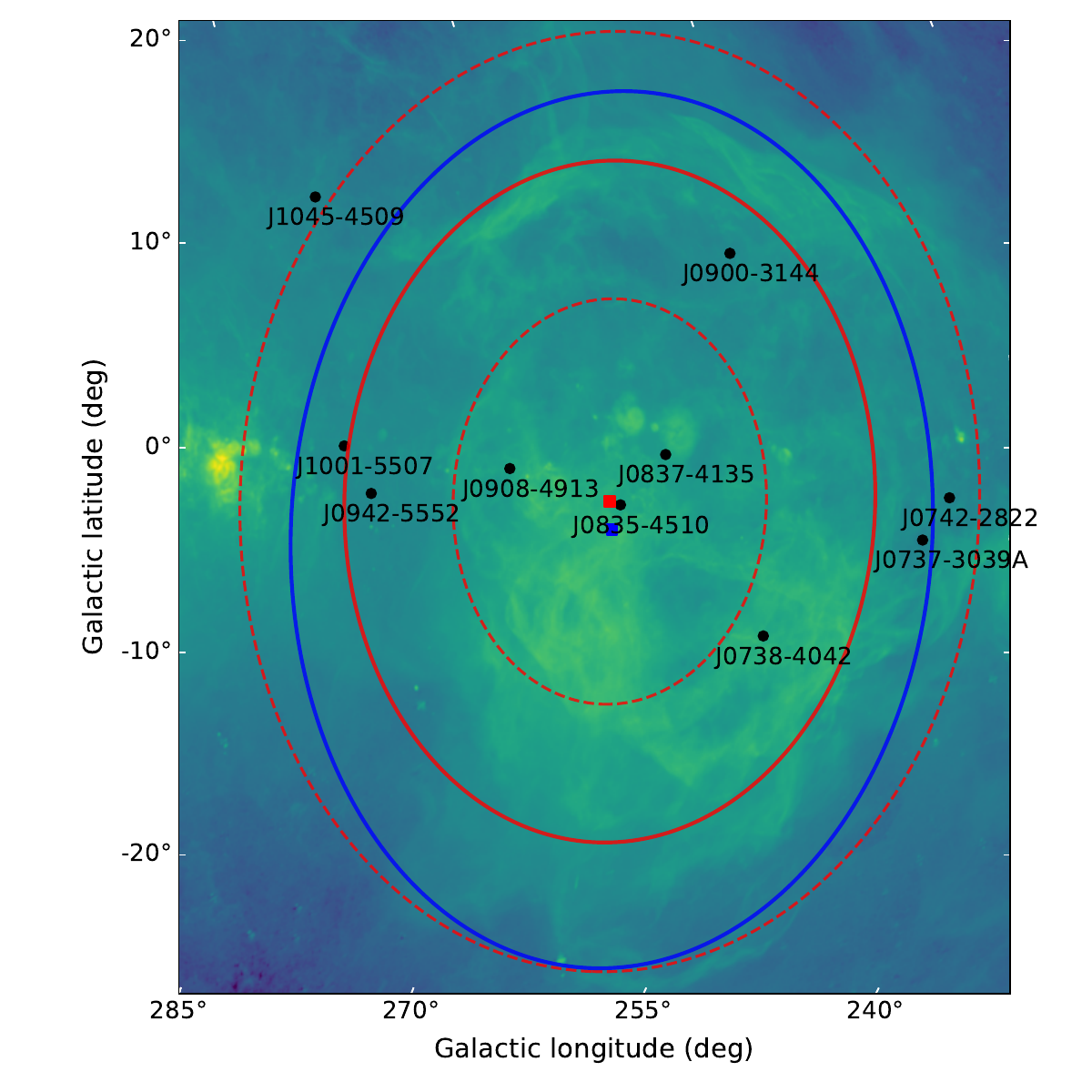}
    \caption{The figure presents the H\,$\alpha$ image of the Gum Nebula overlaid with pulsars in this region, and model ellipses from YMW16 (blue) and our modified nebula model (red), with their respective centres marked by filled squares. The solid red ellipse delineates the peak density in the shell of the GUM25 model, while the dashed red ellipses denote (inner and outer) extents, where electron density described by Gaussian shell profile falls to $1/e$ of the peak density.}
    \label{fig:gum_ellpises}
\end{figure}

On average, the YMW16 model provides more accurate DM-based pulsar distance estimates compared to the NE2001 model (Table \ref{table:psr_dist}). However, surprisingly, the Vela pulsar location is implied to be in front of the Gum Nebula model within YMW16 (Fig. \ref{fig:vela_ne}). The Vela pulsar location has long been discussed, and the broad consensus is that the Vela pulsar is a part of the Gum Nebula, where the nebular shell is likely the dominant scatterer \citep{backer1974, kirsten2019}. The source of such a discrepancy is unclear. Particularly for the Vela pulsar, the YMW16 model underestimates the electron density or overestimates its distance. Recognition of such issues has prompted us to revisit the model and to refine the Gum Nebula model within YMW16 by adjusting all relevant nebular parameters while constraining the $A_{\rm GN}$ using the information on the dominant scatterer for the Vela pulsar (Section \ref{sec:2}) and equation (\ref{eq:ellipse}). We find that the results remain consistent even when the Vela pulsar scatterer-based constraint on $A_{\rm GN}$ is not applied. Additionally, noting the extent of the H\,$\alpha$ map (Fig. \ref{fig:gum_ellpises}), we imposed tighter constraints on key parameters such as $K_{\rm GN}$, $l_c$, and $b_c$, allowing only a narrow range of values during the fitting process. In contrast to the YMW16 model, where certain parameters are fixed, such as the Gum Nebula’s centre ($l_c$, $b_c$, $D_{\rm GN}$) and the scale factor ($K_{\rm GN}$), we allowed all parameters to vary independently within the respective bounds. The parameter bounds used in fitting are as follows: $l_c = (258^\circ, 265^\circ)$, $b_c = (-4^\circ, -1^\circ)$, $n_{\rm GN_0} = (0.1, 2.5)\ {\rm cm}^{-3}$, $W_{\rm GN} = (12, 25)\ {\rm pc}$, $K_{\rm GN} = (1.2, 1.5)$, and $D_{\rm GN} = (300, 500)\ {\rm pc}$. The global optimisation is carried out using the \textit{differential\_evolution} algorithm implemented in the \textsc{SciPy} package. This procedure yields best-fitting parameters of $264.2^\circ$, $-2.63^\circ$, $0.96\,\mathrm{cm^{-3}}$, $23.7\,\mathrm{pc}$, $0.96$, and $341\,\mathrm{pc}$, in that order, with an associated chi-square of eight. To further refine the parameters, we have performed a grid-search optimisation in which the three-dimensional centre of the Gum Nebula and the factor $K_{\rm GN}$ are held fixed to the values obtained above, while the remaining parameters are varied ($A_{\rm GN}$, $W_{\rm GN}$, $n_{\rm GN_0}$). This approach reduces the chi-square from 8 to 5, and the final fitted Gum Nebula parameters are listed in Table \ref{table:gn_components}. The parameters $A_{\rm GN}$ and $W_{\rm GN}$ show significant coupling. A thicker shell with the smaller nebular radius and a thinner shell with larger radius are presently indistinguishable, probably due to the limited sampling of the shell region by the set of the pulsars considered here.

\begin{table*}
    \centering
    \resizebox{\textwidth}{!}{%
    \begin{threeparttable}
    \begin{tabular}{ccccccccccc}
    \toprule
    \addlinespace
    Model & Component & Shape & $W_{\rm GN}$ & $l$ & $b$ & $D_{\text{GN}}$ & $K_{\text{GN}}$ & $A_{\text{GN}}$ & $n_{\rm GN_0}$ & $\chi^2$ \\
    \addlinespace
     & & & (pc) & (deg) & (deg) & (pc) &  & (pc) & (cm$^{-3}$) &  \\
    \midrule
    \addlinespace
    NE2001 &  Gum\,\textsc{i} & Spherical & -- & 260.00 & -1.00 & 500 & 1 & 140 & 0.39 & 209\\
           &  Gum\,\textsc{ii} & Spherical & -- & 262.80 & 2.70 & 500 & 1 & 30 & 0.47 & -- \\
           &  VelaIras & Spherical & -- & 263.25 & -9 & 250 & 1 & 40 & 3.37 & -- \\
           &  Gum\,\textsc{iii} (void) & Spherical & -- & 278.5 & -0.6 & 500 & 1 & 20 & 0.5 & -- \\
    YMW16 & Gum & Ellipsoidal & 15.1 & 264 & -4 & 450 & 1.4 & 125.8 & 1.84 & 28\\
    GUM25 & Gum & Ellipsoidal & 32.6 & 264.2 & -2.63 & 341 & 1.3 & 78.3 & 0.82 & $\sim$5 \\
    \bottomrule
    \end{tabular}
    \caption{Parameters and component(s) of the Gum Nebula model used in NE2001, YMW16, and GUM25 (this work). The nebular parameters include 3-dimensional centre coordinates (columns 5-7), shell thickness (column 4), and size (columns 8-9) of the component(s). The associated electron density of the respective component is listed in column 10. The final column presents the chi-square values, defined based on pulsar distances $\big( \sum_{i=1}^{10} ((D_{\rm measured}^i - D_{\rm model}^i)/\sigma_{\rm measured}^i)^2 \big)$, which are mentioned in Table \ref{table:psr_dist}.}
    \label{table:gn_components}
    \end{threeparttable}
    }
\end{table*}

\begin{table*}
    \centering
    \resizebox{\textwidth}{!}{%
    \begin{threeparttable}
    \begin{tabular}{cccccccccccccccc}
    \toprule
    \addlinespace
    Pulsars & $\tau_{\rm sc}$\tnote{a} & DM\tnote{b} & $D$\tnote{c} & $\langle\zeta\rangle$ & $\langle D_{\rm DT}\rangle$ & $\langle\zeta \times D_{\rm DT}\rangle$ & $D_\text{NE2001}$ & $D_\text{YMW16}$ & $D_\text{GUM25}$ & $\kappa_\nu^{\rm GN}$ & Ref \tnote{d} \\
    \addlinespace
     & ($\mu$s) & (cm$^{-3}$ pc) & (kpc) & & (kpc) & (kpc) & (kpc) & (kpc) & (kpc) & $({\rm m\, pc^{-2}\, cm^6})$ & &\\
    \midrule
    \addlinespace 
    J0737-3039A & -- & 48.9  & 1.1$^{+0.2}_{-0.1}$ \tnote{\textsc{i}} & -- & -- & -- & 0.52 & 1.10 & 1.02 & -- & (\textsc{i, ii}) \\
    \addlinespace
    J0738-4042 & $960 \pm 11$\tnote{\textsc{i}} & 160.5 & $1.6 \pm 0.8$\tnote{\textsc{ii}} & 1.26 & $0.48 \pm 0.13$ & $0.60\pm0.11$ & 2.64 & 1.57 & 1.4 & 27 & (\textsc{i, iii}) \\
    \addlinespace
    J0742-2822 & $8.1 \pm 0.1$\tnote{\textsc{i}} & 73.7  & 2.0$^{+1.0}_{-0.8}$ \tnote{\textsc{ii}} & 0.51 & $4.0\pm0.5$\tnote{$\dagger$} & $1.9\pm0.4$ & 2.07 & 3.11 & 2.86 & 10 & (\textsc{i, iv}) \\
    \addlinespace
    J0835-4510 & $61.9 \pm 0.3$\tnote{\textsc{i}} & 67.8  & $0.28 \pm 0.02$\tnote{\textsc{i}} & 0.79 & $0.33 \pm 0.07$ & $0.26 \pm 0.01$ & 0.24 & 0.33 & 0.27 & 86 & (\textsc{i, v}) \\
    \addlinespace
    J0837-4135 & $22.8 \pm 0.1$\tnote{\textsc{i}} & 147.2 & 1.5$^{+1.2}_{-0.9}$ \tnote{\textsc{ii}} & (0.3, 4) & $(0.3, 1.2)$\tnote{*} & $(0.6, 1.2)$ & 1.04 & 1.42 & 1.63 & 0.84 & (\textsc{i, iii}) \\
    \addlinespace
    J0900-3144 & -- & 75.7  & 0.81$^{+0.38}_{-0.21}$ \tnote{\textsc{i}} & -- & -- & -- & 0.54 & 0.38 & 0.34 & -- & (\textsc{vi}) \\
    \addlinespace
    J0908-4913 & $80 \pm 11$\tnote{\textsc{i}} & 180.4 & 1.0$^{+1.7}_{-0.7}$ \tnote{\textsc{ii}} & 2.09 & $0.28 \pm 0.06$ & $0.58 \pm 0.04$ & 2.57 & 1.03 & 0.89 & 3 & (\textsc{i, iv, viii}) \\
    \addlinespace
    J0942-5552 & $57 \pm 17$\tnote{\textsc{ii}} & 180.2 & 0.3$^{+0.8}_{-0.2}$ \tnote{\textsc{ii}} & (1, 2) & $(0.25, 0.9)$\tnote{\S} & $(0.4, 0.9)$ & 3.77 & 0.41 & 0.43 & 6 & (\textsc{i, iii}) \\
    \addlinespace
    J1001-5507 & $172 \pm 23$\tnote{\textsc{ii}} & 130.3 & 0.3$^{+1.1}_{-0.3}$ \tnote{\textsc{ii}} & 0.92 & $0.41 \pm 0.08$ & $0.38 \pm 0.06$ & 2.78 & 0.41 & 0.36 & 109 & (\textsc{i, iv}) \\
    \addlinespace
    J1045-4509 & -- & 58.1  & 0.34$^{+0.2}_{-0.10}$ \tnote{\textsc{i}} & -- & -- & -- & 1.96 & 0.33 & 0.28 & -- & (\textsc{vii}) \\
    \bottomrule
    \end{tabular}
    \begin{tablenotes}
    \item [a] (\textsc{i}) \cite{krishnakumar2015scatter}, (\textsc{ii}) \cite{mitra2001}.
    \item [b] DM values are taken from PSRCAT \citep{manchester2005australia}
    \item [c] (\textsc{i}) parallax based distance, (\textsc{ii}) kinematic distance.
    \item [d] References for independent distances (column 4); (\textsc{i}) \cite{verbiest2012}, (\textsc{ii}) \cite{deller2009}, (\textsc{iii}) \cite{johnston1996}, (\textsc{iv}) \cite{koribalski1995}, (\textsc{v}) \cite{dodson2003vela}, (\textsc{vi}) \cite{desvignes2016}, (\textsc{vii}) \cite{reardon2015}, (\textsc{viii}) \cite{saravanan1996}.
    \item [$\dagger$] The temporal broadening profile of this pulsar as a function of trial distance exhibits a second solution (Fig.~\ref{fig:tsc_dm_based_dist}), located at $\sim 280$~pc with a higher $\zeta \approx 6.6$.
    \item [*] Values of $\kappa_\nu^{\rm GN} \geq\ 5.6$ are ruled out by the analysis, whereas values $\leq1.2$ have double value solutions (see the text for more details).
    \item [\S] $\kappa_\nu^{\rm GN} \geq 40$ are ruled out (see the text for more details).
    \end{tablenotes}
    \caption{Independent and model-predicted distance estimates of ten pulsars in the direction of the Gum Nebula, which are used in the fitting. The model-predicted distances are shown in columns 8 (NE2001), 9 (YMW16), 10 (GUM25; this work), and 6 and 7 (our new distance estimation technique based on DM and temporal broadening of pulsars, Section \ref{sec:4}). The scaled temporal broadening values using $\alpha=-4$ at 1 GHz are listed in column 2. The average value of $\zeta$, ratio of the measured to the GUM25 model-predicted DM, are listed in column 5. The temporal broadening scaling factors, $\kappa_\nu^{\rm GN}$, at 1 GHz are listed in column 11.}
    \label{table:psr_dist}
    \end{threeparttable}
    }
\end{table*}

\section{Further constraints on pulsar distances}\label{sec:4}
The propagation effects on pulsar signals are implicit distance indicators; pulsar distances can be estimated using related observed quantities, DM, and temporal broadening. As mentioned earlier, it is well known that the estimated pulsar distances using GEDMs can be significantly uncertain if the considered GEDM is poorly constrained in their directions. Independent kinematic distance estimates typically carry large uncertainties and are not feasible along all lines of sight. Therefore, we present a new technique to further refine or constrain the pulsar distances by using simultaneously DM and temporal broadening of pulsars, along with a priori GEDM. This technique iteratively searches for the suitable distance(s) in steps of one parsec, which together satisfy the observed DM and temporal broadening for a given pulsar. For every trial distance ($D_{\rm trial}$), the electron density profile along the line of sight to the pulsar up to $D_{\rm trial}$ is scaled with a factor of $\zeta$, which is the ratio of the measured DM to the model-predicted DM$_{\rm trial}\ (= \int_0^{D_{\rm trial}} n_e dz)$. For a given pulsar, the temporal broadening is then estimated using equation (\ref{eq:temp_broad}), where $\psi(z) = \kappa_\nu (\zeta\times n_e)^2$, assuming $(\Delta n_e)^2 \approx n_e^2$, where $\Delta n_e$ is the electron-density fluctuation along the line of sight. Here $\kappa_\nu$ is a constant factor for a given line of the sight at 1 GHz. For pulsars towards the Gum Nebula, the temporal broadening contribution from the Gum Nebula alone ($\tau_{\rm GN}$) and the rest of the medium ($\tau_{\rm NGN}$) are computed separately using the respective scaled electron density profile ($\zeta \times n_e$) and $\kappa_\nu$ is equal to unity. Then these two contributions are scaled with the respective $\kappa_\nu$ factors, and added to obtain the estimate of the temporal broadening for a given pulsar;
\begin{equation}
    \tau_{\rm sc}(D_{\rm trial}) = \kappa_\nu^{\rm NGN} \times \tau_{\rm NGN}\ +\ \kappa_\nu^{\rm GN} \times \tau_{\rm GN}.
\end{equation}

\begin{figure}
    \centering
    \includegraphics[width=1\linewidth]{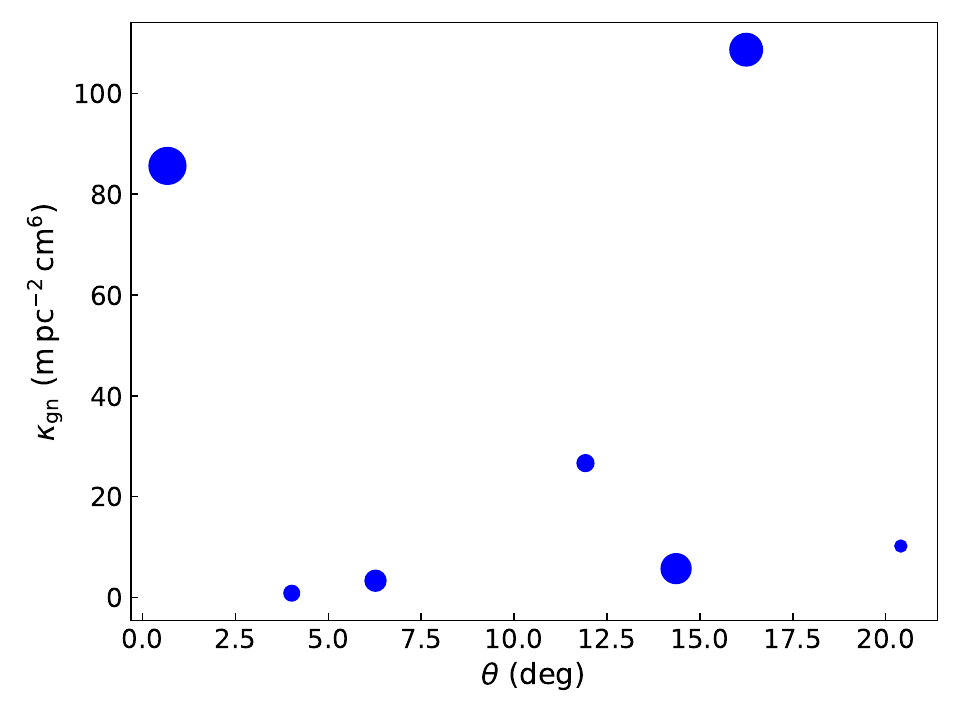}
    \caption{The plot shows the variation of the $\kappa_\nu^{\rm GN}$ factor with angular separation of pulsars from the centre of the Gum Nebula. The dot sizes are scaled inversely with distance, such that nearby pulsars appear larger. It is clearly seen that the $\kappa_\nu$ factor values vary significantly across different lines of sight through the nebula.}
    \label{fig:k_factor}
\end{figure}

Such a linear combination of $\tau_{\rm sc}$ is justified, noting its underlying dependence on the square of the scattering angle, which is to be summed incoherently.
For the non-Gum case, the $\kappa_\nu$ value is anchored on PSR J0820$-$1350, which lies outside the Gum Nebula and has no contribution from the other Galactic components \citep{yao2017new}. Its distance \citep{chatterjee2009} and temporal broadening measurements \citep{kl07} are well determined, yielding $\kappa_{\rm NGN} = 1.02 \pm 0.52\,{\rm m\,pc^{-2}\,cm^6}$. 
Now, we need a similar estimate, $\kappa_\nu^{\rm GN}$, for the nebula. As seen from the Table \ref{table:psr_dist} or Fig. \ref{fig:k_factor}, the $\kappa_\nu^{\rm GN}$ values estimated assuming distance given by $D_{\rm GUM25}$, vary over a wide range. Instead of using a single value of $\kappa_\nu^{\rm GN}$, we obtain a conservative estimate of the distance ($\langle D_{\rm DT}\rangle$) by averaging distances suggested by this entire range of $\kappa_\nu^{\rm GN}$ (from 0.84 to 109 in steps of five). The uncertainty in $\langle D_{\rm DT}\rangle$ combines the root-mean-square (RMS) variation in $D_{\rm DT}$ computed using different $\kappa_\nu^{\rm GN}$ values, and that resulting from the uncertainties in $\tau_{\rm sc}$. For all pulsars in our sample, except J0942-5552 (where fractional temporal broadening uncertainty is large), the dominant uncertainty contribution come from the RMS variation in $D_{\rm DT}$.

\begin{figure*}
    \centering
    \includegraphics[width=0.49\linewidth]{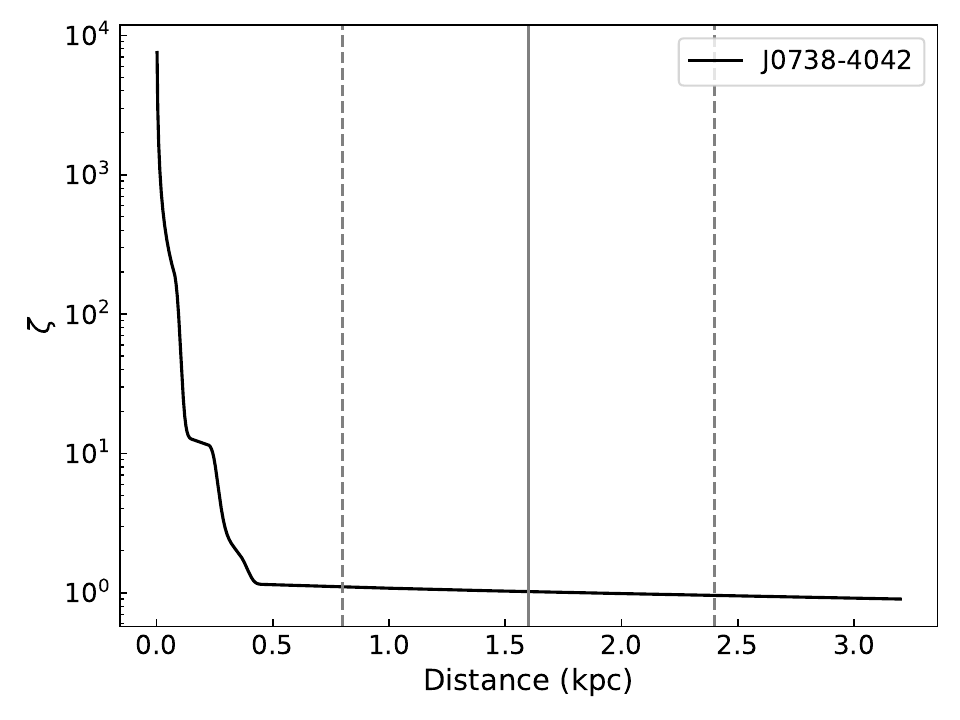}
    \includegraphics[width=0.49\linewidth]{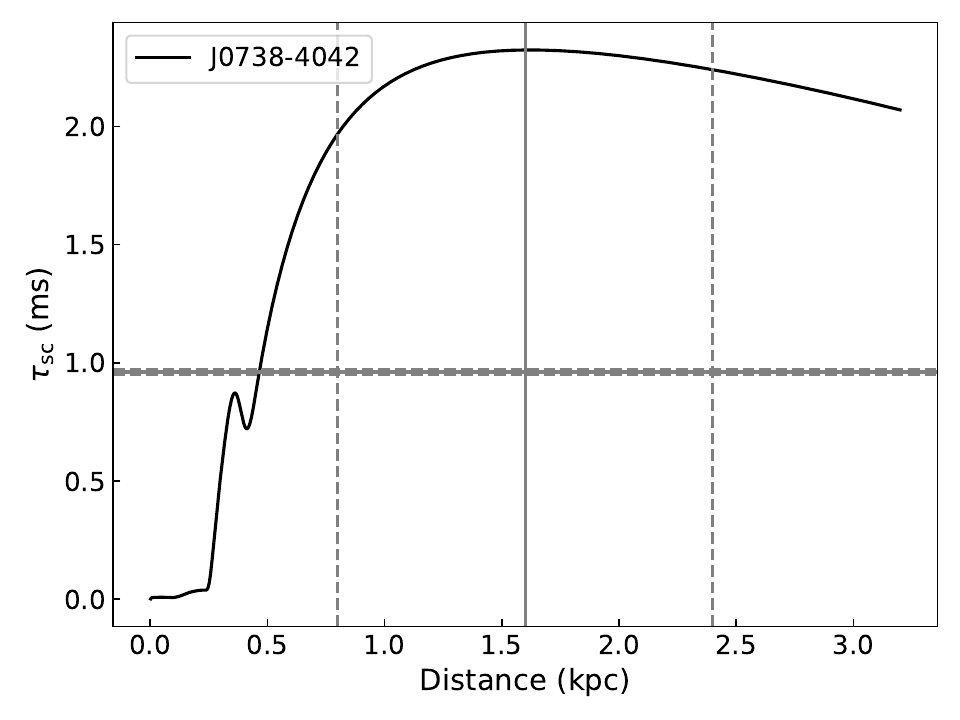}
    \includegraphics[width=0.49\linewidth]{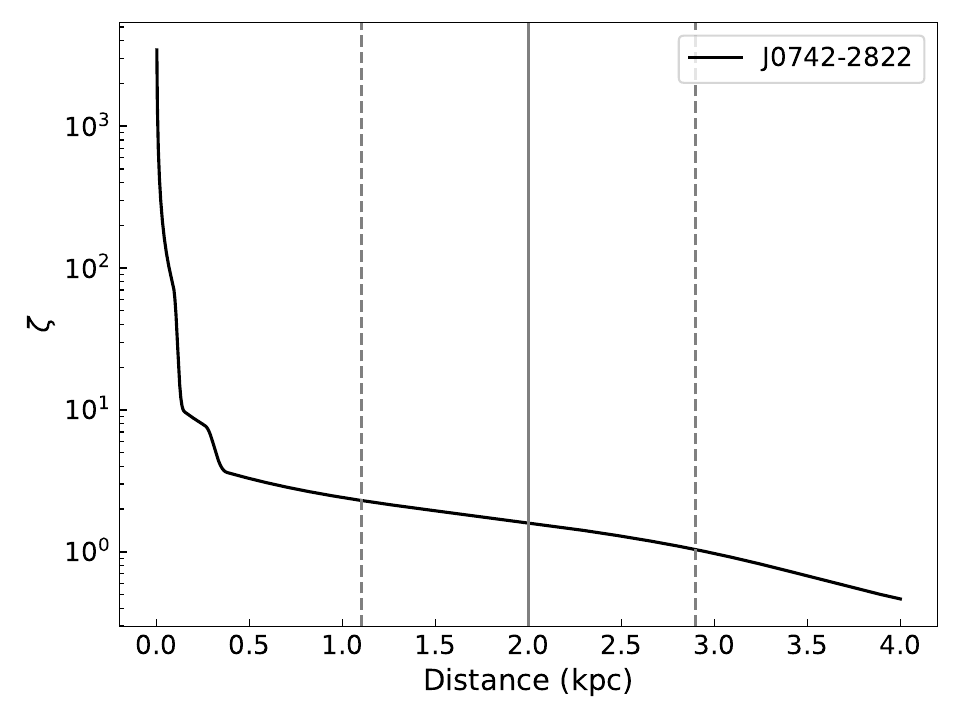}
    \includegraphics[width=0.49\linewidth]{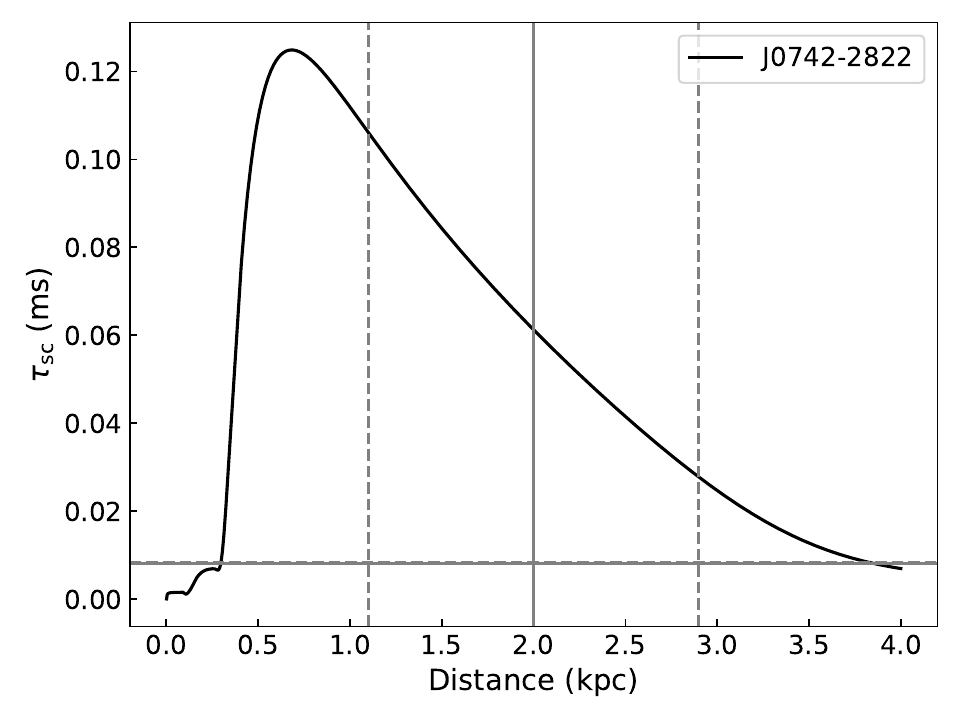}
    \includegraphics[width=0.49\linewidth]{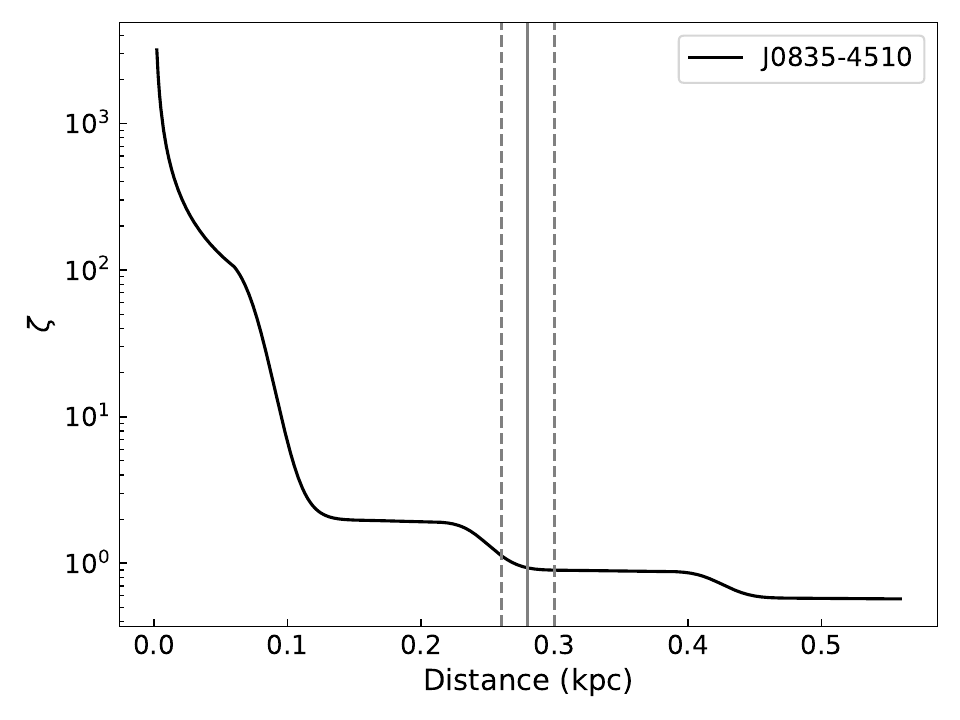}
    \includegraphics[width=0.49\linewidth]{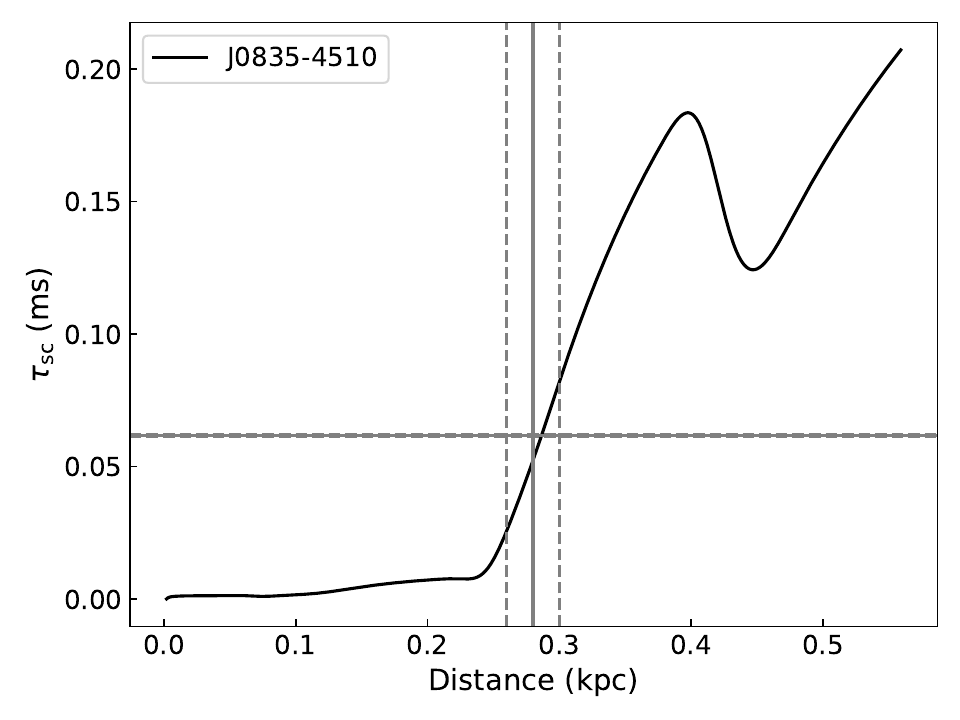}
    \caption{Illustration of our new technique for pulsar distance estimation using examples of J0738-4042, J0742-2822, and J0835-4510. The horizontal and vertical solid and dashed lines mark the respective independently measured quantity and their uncertainties, respectively. The uncertainties in temporal broadening values are relatively much smaller, and hence, horizontal lines almost overlap. The measured temporal broadening and associated uncertainties are scaled to 1 GHz using $\alpha = -4$. The $\tau_{\rm sc}$ plots clearly show the oscillations in the estimated temporal broadening values at the electron density enhanced regions; this feature can not be captured in the DM$^\beta$ modelling. In some cases, multiple distance solutions are possible, but can be discarded based on the $\zeta$ factor value, which should be close to unity. Here, $\kappa_\nu^{\rm GN}$ is assumed to be $\sim 50\, {\rm m\, pc^{-2}\, cm^6}$.}
    \label{fig:tsc_dm_based_dist}
\end{figure*}

For each $\kappa_\nu^{\rm GN}$, we estimate the optimal trial distance by assessing that the relative difference between the measured and modelled temporal broadening, $\left( \tau_{\rm sc}^{\rm measured} - \tau_{\rm sc}^{\rm model} \right) / \tau_{\rm sc}^{\rm measured}$, is less than 5\%. This tolerance accounts for the finite sampling of both the distance and $\kappa_\nu^{\rm GN}$ grids and avoids the need for excessively fine grids that would increase the computational cost. For a given pulsar, some (particularly lower) values of $\kappa_\nu^{\rm GN}$ are not adequate to match the modelled temporal broadening with the measured value, and therefore, the mean and RMS uncertainty of the optimal distances are computed over only acceptable solutions (Table \ref{table:psr_dist}). The new technique might appear to perform poorly for J0837–4135 because the measured broadening is unusually low, and therefore the $\kappa_\nu^{\rm GN} \geq 5.7$ are not allowed because the non-Gum contribution to temporal broadening alone exceeds the measured value at some trial distance. A similar situation arises for J0942–5552, for which $\kappa_\nu^{\rm GN} \geq 40$ are not permitted. For these two pulsars we have indicated the allowed range of distances instead of a specific distance with uncertainty estimate. The pulsar J0942-5552 has double distance solutions (Table~\ref{table:psr_dist} and Fig.~\ref{fig:tsc_dm_based_dist}). Additionally, the distance to J0742–2822 is highly sensitive to the adopted Gum Nebula parameters, a dependence that is not fully reflected in the quoted uncertainties. Therefore, for pulsars located near the nebular boundary, the resulting uncertainty estimates are not very robust.
Future VLBI observations to estimate parallax-based distance for this pulsar would indeed of much significance for the Gum nebula modelling.

In the temporal broadening formalism, $\zeta$ and the trial distance are coupled: smaller trial distances understandably yield larger $\zeta$ values, and vice versa (Fig. \ref{fig:tsc_dm_based_dist}). Consequently, the product $\zeta \times D_{\rm trial}$ serves as a more appropriate distance estimate. This can be appreciated readily by considering the case of the uniform electron density. In this illustrative case, all trial distances along with an appropriate single value of $\kappa_\nu$ factor can satisfy requirement of matching the observed temporal broadening.
Hence, the average distance estimate will be dictated by the range of trial distances. In contrast, the product $\zeta \times D_{\rm trial}$ would be same for all trial distances, and the average will not have any undesired dependence on the range of trial distances examined. In a general case, the mean density square times distance square term in the integral for $\tau_{\rm sc}$ relates to the ${\rm DM}^2$ dependence, and it is the variation in density and $\kappa_\nu$ along the sight-line which contribute to departure from a mere ${\rm DM}^2$ dependence. At very short trial distances (less than 0.5 kpc), the true density may not follow the long distance mean in GEDM, and can cause significant uncertainty in distance estimates, if the variance of changing density also deviates from that over longer distance scales. However, using the $\zeta \times D_{\rm trial}$ product as a distance estimate and averaging the optimal distances  (disregarding outliers, when appropriate) obtained for a given/assumed distribution of $\kappa_\nu$ mitigates many of these issues. 
The optimum trial distance will be equal to the true distance, as well as $\zeta \times D_{\rm trial}$ product  when the considered GEDM adequately describes the electron density along the line of sight, thus when $\zeta=1$.
The product of $\zeta$ and distance are also provided in Table \ref{table:psr_dist}, and we consider them as more appropriate distance estimates.

As discussed above, we have explored a possible range of $\kappa_\nu^{\rm GN}$ values for these directions. Additionally, we highlight that the predicated temporal broadening (as a function of $D_{\rm trial}$) does not increase monotonically with increasing $D_{\rm trial}$, but nonetheless interestingly, oscillates around the enhanced electron density region, like local bubble, Gum Nebula, etc. (Fig. \ref{fig:tsc_dm_based_dist}). Hence, our approach of solving for distance employing the full integration to estimate the temporal broadening (equation \ref{eq:temp_broad}) is more robust and offers an effective way to obtain better constraints on pulsar distances.

\section{Discussion and Conclusion}\label{sec:5}
The distance estimation method for pulsars, introduced by \cite{deshpande1998improving}, is employed here to estimate the distance to the dominant scattering screen in the sight line to the pulsar with known distance. This technique provides the fractional distance to either the pulsar or the scatterer by utilising the pulsar's scintillation parameters and proper motion. The fractional distance can then be converted to a physical distance if an independent distance estimate for either the pulsar or the scatterer is available. Some care needs to be taken when using pulsar scintillation parameters, as they should be derived from sufficiently long temporal baseline observations. Otherwise, the estimates may be biased by refractive scintillation \citep{romani1986refractive}, the Earth’s orbital motion, or the binary motion of the pulsar (if applicable). We have used an inverted version of this method to estimate the frontal edge of the Gum Nebula, which is turn out to be $196 \pm 16$ pc from its centre, for the given distance of the Vela pulsar. This translates to the angular radius of the Gum Nebula to be $24^\circ \pm 2^\circ$, assuming the nebula is spherical and at a distance of 450 pc. Our estimate of the nebula radius is within the error bar of the angular radius estimated by \cite{purcell2015radio} under the same assumptions. Our approach provides a robust estimate of the nebula's frontal edge and can be extended to other such systems. Additionally, the method also provides a scattering strength factor for the thin screen, which in our case is around 384, comparable to the theoretical prediction, assuming $\Delta n_e/n_e \sim 1$, ($n_{GN_0}^2 \times(2W_{\rm GN}) / n_{\rm uniform}^2 \times D$) which turns out $\sim 150$, for a uniform electron density to be 0.03 cm$^{-3}$. Since this method relies on a two-component scattering model, it does not provide direct insights into the structure of the Local Bubble (Fig. \ref{fig:vela_ne}) in our case. However, the consistency of our results with previous studies suggests that the Local Bubble has little to no significant impact on the scattering of the Vela pulsar, and its associated electron density is likely to be overestimated in YMW16 compared to the estimate by \cite{bhat1998}.

In the directions of the Gum Nebula region, nearly 100 pulsars are known, yet only four have reliable parallax measurements, and a total of ten have independent distance estimates (Table \ref{table:psr_dist}), leading to less accurate GEDMs in this region. Based on the available data on pulsars, we have assessed that the YMW16 model generally performs better in estimating pulsar distances in this region, while the NE2001 model provides better temporal broadening predictions. This difference likely arises because the temporal broadening values for pulsars were used to constrain the NE2001 model but not to constrain the YMW16 model. However, these two models are still far from matching the observed temporal broadening. We have further refined the nebula parameters within the YMW16 model based on distances of ten pulsars to provide even more consistent distances. Notably, an important outcome is that now the GNED model (GUM25) suggests the Vela pulsar is behind the Gum Nebula frontal shell, rather than outside (Fig. \ref{fig:vela_ne}), and also provides more accurate distances for the pulsars in this region. Further refinements are expected as more independent pulsar distances become available. Notably, only three pulsars with known distances are located above and below the Galactic plane, but incorporating more distance estimates from this population, as they become available, will impose tighter constraints on the $K_{\rm GN}$ factor.

Our new proposed method for distance estimation based on joint DM and temporal broadening of pulsars is more robust than DM-based distance estimation since it first refines the electron density profile using DM value and then uses appropriately weighted integral form of temporal broadening to estimate the distance. The technique also employs a factor, $\kappa_\nu$, which may vary for different regions of the ISM but can be assumed to be approximately the same for closely spaced lines of sight. Notably, this approach also suggests that the DM$^\beta$ ($\beta$>1) is a poor estimator for the temporal broadening, since in reality, scatter broadening also depends on the location of the scatterer along with the electron density fluctuations. 

In principle, the new distance estimation technique can be used to develop and refine GEDMs in their smooth components, as well as providing indications that may justify addition of certain details, such as clumps and voids along specific lines of sight \citep{cordes_ne2001}, guided by the values of $\zeta$ and $\kappa_\nu$. We have applied this approach to modify both the Gum and non-Gum Nebula components of the electron density model. Unlike Section \ref{sec:3}, where only independent distances were considered to assess DM-based distances implied by density model, we can incorporate the $\zeta.D_{\rm DT}$ distances, which provide improved distance constraints, and iteratively update the Gum Nebula parameters, allowing both $\kappa_\nu^{\rm GN}$ and $\kappa_\nu^{\rm NGN}$ to vary freely. Our initial tests towards this have shown encouraging indications. However, in light of the wide range of variation of $\kappa_\nu^{\rm GN}$ and the very limited sample pulsars we have (let alone those with reliable independent distances), which together render inadequate degrees of freedom, it is difficult to quantify such further refinement in the present case. 

It is clear that the Gum Nebula is a complex structure that includes the IRAS Vela Shell, cometary globules, and other features, as well as Vela pulsar's own wind nebula. The Local Bubble should also be treated separately, as these components and substructures may have distinct $\kappa_\nu$ factors, and by explicitly accounting for them, a more accurate GEDM could be constructed. Although our new distance technique has been described and employed here in a specific context, it is applicable in general in any other sector. With an adequate sample of pulsars with scattering measurements, a reasonable initial electron density model, and suitable constraints on $\kappa$, the $\zeta.D_{\rm DT}$ averages can provide distance estimates comparable to or better than kinematic distances at the least, and a scope for refinement of density models with assessment of convergence via internal consistency (such as narrow spread in $\kappa$ and $\zeta$ approach to unity). Even in the context of presently available models, such as YMW16, our method will aid in estimating pulsar distances in directions where kinematic methods are unreliable or not possible, provide tighter distance constraints, and ultimately improve Galactic electron density models.

\section*{Acknowledgements}
We thank the anonymous referee for the detailed review, and for insightful comments and suggestions, which significantly improved the manuscript. This work made use of several Python packages, including numpy \citep{Harris20}, scipy \citep{2020SciPy-NMeth}, astropy \citep{Astropy-Collaboration13, Astropy-Collaboration18, Astropy-Collaboration22}, matplotlib \citep{Hunter:2007}, PyGEDM \citep{Price_Flynn_Deller_2021}, Pandas \citep{reback2020pandas}, and psrqpy \citep{pitkin2018_psrqpy}.

\section*{Data Availability}
The data underlying this paper and codes will be shared on reasonable request to the corresponding author. The H\,$\alpha$ data are publicly available on \url{https://skyview.gsfc.nasa.gov/current/cgi/query.pl}.

\bibliographystyle{mnras}
\bibliography{references} 



\bsp	
\label{lastpage}
\end{document}